\documentclass[twocolumn,showpacs,preprintnumbers,aps]{revtex4}

\newcommand{\BE}{\begin{equation}}
\newcommand{\EE}{\end{equation}}
\newcommand{\BA}{\begin{eqnarray}}
\newcommand{\EA}{\end{eqnarray}}

\begin{document}
\preprint{0202}

\title{Water wave propagation and scattering over topographical bottoms}
\author{Zhen Ye}\email{zhen@phy.ncu.edu.tw}\affiliation{Department
of Physics, National Central University, Chungli, Taiwan 32054}
\date{\today}

\begin{abstract}

Here I present a general formulation of water wave propagation and
scattering over topographical bottoms. A simple equation is found
and is compared with existing theories. As an application, the
theory is extended to the case of water waves in a column with
many cylindrical steps.

\end{abstract}


\maketitle

\section{Introduction}

There have been many approaches for investigating propagation of
water waves over various bottom topographies. A vast body of
literature exists. For brevity, I refer the reader to the textbook
\cite{CCM}. Here I would like to derive from the first principle a
simple but coherent formulation for the problem. It will be shown
that this approximate approach compares favorably with existing
approximations when applied to the cases considered previously.
The advantage of the present approach is obvious: it is simple,
accommodating, systematic, and can be easily numerically
programmed. In particular, here I explicitly show that it
respectively recovers three previous results for shallow water,
deep water, and scattering by rigid cylinders standing in water. I
will first give a theory for general bottom topographies. Then I
will extend to study the case of water wave propagation and
scattering in a column with many cylindrical steps.

\section{General theory}

Consider a water column with an arbitrary bottom topography. We
set up the coordinates as follows. Let the $z$ axis be vertical
and directed upward. The $x-y$ plane rests at the water surface
when it is calm. The depth of the bottom, which describes the
bottom topography, is denoted by $h(x, y)$, and the vertical
displacement of the water surface is $\eta(x, y, t)$. Now we
derive the governing equations for the water waves.

Consider a vertical column with a base differential element $dxdy$
at $(x,y)$. The change rate of the the volume of the column is
$$\frac{\partial}{\partial t}\eta (x,y,t)dxdy.$$ By conservation
of mass, this would equal to the net volume flux from all the
horizontal directions, i.~e. $$ \frac{\partial}{\partial t}\eta
(x,y,t)dxdy = - \nabla_\perp\cdot \left[\int_{-h}^\eta
dz\vec{v}_\perp(x,y, z, t)\right]dxdy,$$ where $\nabla_\perp =
(\partial_x,
\partial_y)$, and `$\perp$' denotes the horizontal directions.
This gives us the first equation \BE \frac{\partial}{\partial
t}\eta (x,y,t) = - \nabla_\perp\cdot \left[\int_{-h}^\eta
dz\vec{v}_\perp(x,y, z, t)\right] \label{eq:1a} \EE

The second equation is obtained from the Newton's second law. From
the Euler equation for incompressible ideal flows
$$
\partial_t\vec{v} + (\vec{v}\cdot\nabla)\vec{v} =
-\frac{1}{\rho}\nabla p - \rho g \hat{z},$$ which is valid at
$z=0$, with $g$ being the gravity acceleration, and \BE p = \rho
g(\eta-z), \label{eq:press}\EE we obtain \BE \label{eq:2a}
\frac{\partial }{\partial t} \vec{v}_\perp(x, y, 0, t) +
\left[(\vec{v}\cdot\nabla)\vec{v}\right]_{\perp, z=0} =
-g\nabla_\perp\eta.\EE Note when the liquid surface tension is
included, the following term should be added to
Eq.~(\ref{eq:press}) \BE \sigma\nabla_\perp^2 \eta,\EE in which
$\sigma$ is the surface tension coefficient. In this paper, for
short, we ignore this effect.

Another equation is from the boundary condition at $z=h$, which
states \BE \left.\vec{v}\cdot\hat{n}\right|_{z=-h(x,y)} =
0,\label{eq:3a}\EE where $\hat{n}$ is a normal to the bottom. For
an incompressible fluid, we also have the following Laplace
equation, \BE \nabla\cdot\vec{v}(x,y,z,t) = 0, \label{eq:4a} \EE
in the water column.

Equations (\ref{eq:1a}), (\ref{eq:2a}), (\ref{eq:3a}), and
(\ref{eq:4a}) are the four fundamental equations for water waves.

\subsection{Linearization}

For small amplitude waves, i.~e. $\eta << h$, we can ignore the
non-linear terms in (\ref{eq:1a}) and (\ref{eq:2a}). Such a
linearization leads to the following two equations \BE
\frac{\partial}{\partial t}\eta (x,y,t) = - \nabla_\perp\cdot
\left[\int_{-h(x,y)}^0 dz\vec{v}_\perp(x,y, z, t)\right],
\label{eq:b} \EE and \BE \frac{\partial}{\partial t}
\vec{v}_\perp(x,y,0,t) + g\nabla_\perp\eta(x,y,t) = 0.
\label{eq:c}\EE These two equations together with
Eqs.~(\ref{eq:3a}) and (\ref{eq:4a}) determines scattering of
water waves with a bottom topography.

\subsection{Propagation approximation}

Here we provide an approximate solution to Eqs.~(\ref{eq:b}),
(\ref{eq:c}), (\ref{eq:3a}), and (\ref{eq:4a}). The procedure is
as follows. When the variation of the bottom topography is smaller
than the wavelength (to be determined self-consistently), we can
first ignore terms involving $\nabla_\perp h$, and solve for the
velocity field. For the incompressible fluid, the velocity field
can be represented by a scalar field, i.~e.
$$\vec{v}(x,y,z,t) = \nabla\Phi(x,y,z,t).$$ We write
all dynamical variables with a time dependence $e^{-i\omega t}$
(this time fact is dropped afterwards for convenience). This
procedure leads to the following equations for $\Phi$. \BE
\nabla^2\Phi(x, y , z)=0. \label{eq:phi1}\EE with \BE
\omega^2\Phi(\vec{r}, 0) + g \frac{\partial}{\partial z}
\Phi(\vec{r}, 0) = 0; \ (\vec{r} = (x,y)). \label{eq:phi2}\EE The
first approximation is made at the bottom ($z=-h$). The boundary
condition at the bottom reads \BE \frac{\partial}{\partial
n}\Phi(\vec{r},-h) = \frac{\partial }{\partial z} \Phi(\vec{r},
-h) + \nabla_\perp\cdot\nabla_\perp \Phi(\vec{r},-h) = 0.\EE We
approximate that $\hat{n}$ is in the $z$ direction by neglecting
the second term in the above equation. This is valid as long as
$\nabla_\perp h << kh$. Thus the boundary condition gives \BE
\frac{\partial }{\partial z} \Phi(\vec{r}, -h) = 0.
\label{eq:phi3}\EE Note that this condition is exact in the case
of step-wise topographical bottoms, to be discussed later.
Eqs.~(\ref{eq:phi1}), (\ref{eq:phi2}), and (\ref{eq:phi3}) lead to
the solution for $\Phi$ \BA \Phi(x,y,z) &=&
\phi(x,y) \cosh(k(z+h)) + \nonumber\\
& & \sum_{n} \phi_n(x,y) \cos (k_n(z+h)),\label{eq:phi4}\EA where
$k$ satisfies \BE \omega^2 = gk(x,y)\tanh(k(x,y)h(x,y)), \EE and
$k_n$ satisfies \BE \omega^2 = gk_n(x,y)\tan(k_n(x,y)h(x,y)).\EE
Here $\phi$ and $\phi_n$ are determined by \BE (\nabla^2_\perp +
k^2)\phi =0, \EE and \BE (\nabla^2_\perp - k_n^2)\phi_n =0.
\label{eq:eve}\EE Eq.~(\ref{eq:eve}) leads to evanescent wave
solutions.

The second approximation is to ignore the summation terms in
Eq.~(\ref{eq:phi4}). Such an approximation is based upon the
following consideration. The summation terms represent the
correction of evanescent waves caused by irregularities such as
sudden changes of depth. As these waves are spatially confined, it
is reasonable to expect that such a correction will not affect the
overall wave propagation, and the general features of the wave
propagation. Indeed, when we apply the later approximate solution
to the extreme case of propagation of water waves over an infinite
step, we find that our results agree reasonably well with that
from two other approximate approaches\cite{Newman,Miles}. For
example, the difference in the reflection results is uniformly
less than a few percent for a wide range of frequencies. The
largest discrepancy can happen for the transmission results, but
the difference is still less than 15\%. Furthermore, we find that
the derived result is in agreement with that of Kirby for the case
of waves over a flat bed with small ripples \cite{Kirby}. As
matter of fact, in this case, it can be shown that after a
mathematical manipulation\cite{math} Eq.~(2.11) in \cite{Kirby}
becomes essentially the same as the following
Eq.~(\ref{eq:finala}).

\input epsf.tex
\begin{figure}
\epsfxsize=2.5in\epsffile{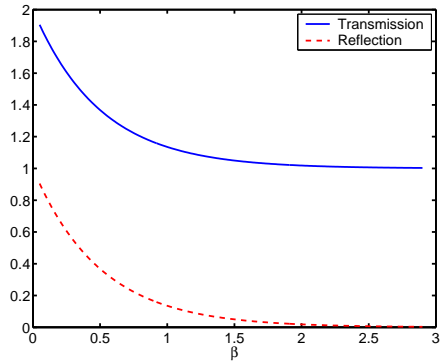} \smallskip
\caption{Transmission and reflection coefficients versus
$\beta=kh$ for an infinite step, obtained from
Eq.~(\ref{eq:finalb}). While the result for the reflection agrees
very well with that in Refs.~\cite{Newman,Miles}, there is some
discrepancy in the transmission results within the range of $kh$
between 0.4 to 1.2; the largest discrepancy of about 15\% occurs
around $kh=0.8$ for the transmission. The legends are adopted from
\cite{Miles} } \label{fig2}
\end{figure}

Under the above approximations, we have \BE \Phi(x,y,z) \approx
\phi(x,y) \cosh(k(z+h)), \label{eq:phi4a}\EE and \BE
v_\perp(x,y,z) \approx \cosh(k(z+h))\nabla_\perp\phi .
\label{eq:phi4b}\EE Now taking Eqs.~(\ref{eq:phi4a}) and
(\ref{eq:phi4b}) into Eqs.~(\ref{eq:b}) and (\ref{eq:c}), we get
\BE \nabla_\perp\left(\frac{\tanh(kh)}{k}\nabla_\perp\eta\right) +
\frac{\omega^2}{g}\eta = 0.\EE For convenience, hereafter we write
$\nabla_\perp$ as $\nabla$ when it acts on the surface wave field
$\eta$. That is \BE
\nabla\left(\frac{\tanh(kh)}{k}\nabla\eta\right) +
\frac{\omega^2}{g}\eta = 0,  \label{eq:finala}\EE or \BE
\nabla\left(\frac{1}{k^2}\nabla\eta\right) + \eta = 0,
\label{eq:finalb}\EE where $k$ satisfies \BE \omega^2 =
gk(\vec{r})\tanh(k(\vec{r})h(\vec{r})). \label{eq:dispersion}\EE
From this equation, we can have the conditions linking domains
with different depths as follows: both $\eta$ and
$\frac{\tanh(kh)}{k}\eta = \frac{\omega^2}{gk^2}\eta$ are
continuous across the boundary.


Eq.~(\ref{eq:finalb}) is similar to what is known as the mild-slop
approximation\cite{CCM}: \BE
\frac{1}{c}\nabla\left(\frac{c}{k^2}\nabla\eta\right) + \eta = 0,
\label{eq:finald}\EE where $c$ is given by \BE c =
\frac{1}{2}\left(1 + \frac{2kh}{\sinh(2kh)}\right). \EE
Eq.~(\ref{eq:finald}) was derived by a number of authors under the
situation that $\nabla h << kh$. In fact, under this condition it
can be shown that Eq.~(\ref{eq:finalb}) and Eq.~(\ref{eq:finald})
are equivalent.


Note that when the surface tension is added, Eq.~(\ref{eq:finala})
becomes \BE \nabla\cdot\left[\frac{\tanh(kh)}{k}\nabla\left(\eta -
\frac{\sigma}{g\rho}\nabla^2\eta \right) \right] +
\frac{\omega^2}{g}\eta = 0, \label{eq:finalc} \EE with
Eq.~(\ref{eq:dispersion}) becoming \BE \omega^2 = \left(gk +
\frac{\sigma}{\rho}k^3\right)\tanh(kh).\EE

\subsection{The situation of shallow water or low frequencies}

In the case of shallow water, i.~e. $kh <<1$, we obtain from
Eq.~(\ref{eq:finala}) \BE \nabla\cdot(h\nabla\eta) +
\frac{\omega^2}{g} \eta = 0. \label{eq:final01}\EE This is the
fundamental equation governing the small amplitude waves in
shallow water, first derived by Lamb\cite{Lamb}.

\subsection{The situation of deep water or high frequencies}
For the deep water case, $hk >>1$, we have \BE k =
\frac{\omega^2}{g}, \label{eq:deep1}\EE and \BE \nabla^2\eta +
\frac{\omega^4}{g^2}\eta=0.\label{eq:deep2}\EE In the deep water,
the dispersion relation is not affected by the bottom topography.

\subsection{Scattering by infinite rigid cylinders}

Equations (\ref{eq:finala}) or (\ref{eq:finalb}) are also
applicable to another class of situation which has been widely
studied in the literature. That is, the scattering of water waves
by infinite rigid cylinders situated in a uniform water column.
When applying (\ref{eq:finala}) or (\ref{eq:finalb}) to this case,
we find that these two equations are actually exact. In the
medium, the wave equation is \BE (\nabla+k^2)\eta = 0\EE with the
boundary condition at the $i$-th cylinder\BE
\left.\hat{n}_i\cdot\nabla\eta\right|_{i} = 0,\EE obtained as we
set the depths of the cylinders equal zero; $\hat{n}_i$ is a
normal to the interface. In fact, in this case, the problem
becomes equivalent to that of acoustic scattering by rigid
cylinders, and all the previous acoustic results will
follow\cite{Sanchez,1998,Chen,Chen1}, such as the interesting
phenomenon of deaf bands.

\section{Water waves in a water column with cylindrical
steps}

\begin{figure}
\epsfxsize=2.5in\epsffile{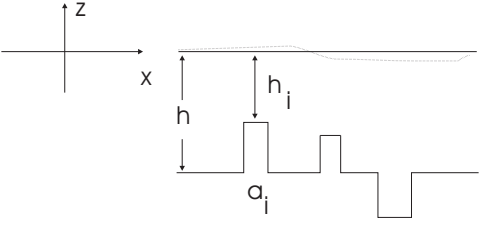} \smallskip
\caption{Conceptual layout (side view of the three dimensional
coordinates): There are $N$ cylindrical steps located in a water
column with depth $h$. The depths of the steps are denoted by $h_i
(i =1,2,\dots, N)$ measured from the upper surface of the water
column, and the radii of the steps are denoted by $a_i$. The
coordinates are set up as shown. The steps are located at
$\vec{r}_i$. The $y$-axis lies perpendicularly to the page}
\label{fig1}
\end{figure}

The problem we are now going to consider is illustrated by
Fig.~\ref{fig1}. We consider a water column with a uniform depth
$h$. There are $N$ cylindrical steps (or holes when $h_i>h$)
located in the water. The depths of the steps are measured from
the water surface and are denoted by $h_i$ and the radii are
$a_i$. In the realm of the linear wave theory, we study the water
wave propagation and scattering by these steps.

\input epsf.tex

\subsection{Band structure calculation}

When all the steps are with same $h_1=h_2=\cdots=h_N$ and the
radius $a$, and are located periodically on the bottom, then we
can use Bloch's theorem to study the water wave propagation.
Assume the steps are arranged either in the square or hexagonal
lattices, with lattice constant $d$. Here we use the standard
plane-wave approach\cite{Kush,Msc}. By Bloch's theorem, we can
express the field $\eta$ in the following form \BE \eta(x,y) =
e^{i\vec{K}\cdot\vec{r}} \sum_{\vec{G}} C(\vec{G},
\vec{K})e^{i\vec{G}\cdot\vec{r}},\label{eq:eta}\EE where $\vec{r}
= (x,y)$, $\vec{G}$ is the vector in the reciprocal lattice, and
$\vec{K}$ the Bloch vector.

In the present setup, the bottom topograph is periodic, so we have
the following expansion \BE \frac{\tanh(kh)}{k} = \sum_{\vec{G}}
A(\vec{G})e^{i\vec{G}\cdot\vec{r}},\label{eq:h}\EE with \BE
A(\vec{G}) =  \left(\frac{\tanh(k_1h_1)}{k_1}-
\frac{\tanh(kh)}{k}\right) f_s + \frac{\tanh(kh)}{k},\EE
for $$
\vec{G}=0;$$ and \BE A(\vec{G}) =
 \left(\frac{\tanh(k_1h_1)}{k_1} -
\frac{\tanh(kh)}{k}\right)F_s(\vec{G}),\EE for $$ \vec{G} \neq
0.$$ Here $k_1$ and $k$ are determined by \BE \omega^2 =
gk_1\tanh(k_1h_1) = gk\tanh(kh),\EE and $f_s$ is the filling
factor given by\cite{Msc}
$$f_s =  \left\{\begin{array}{ll} \pi\left(\frac{a}{d}\right)^2, & \mbox{Square lattice}\\
\frac{2\pi}{\sqrt{3}}\left(\frac{a}{d}\right)^2, & \mbox{Hexagonal
lattice}, \end{array}\right.
$$ and $F_s$ is the structure
factor
$$F_s(\vec{G}) = 2f_s\frac{J_1(|\vec{G}|a)}{|\vec{G}|a}.$$

Substituting Eqs.~(\ref{eq:eta}) and (\ref{eq:h}) into
Eq.~(\ref{eq:finala}), we get \BE
\sum_{\vec{G}'}Q_{\vec{G},\vec{G}'}(\vec{K},\omega)C(\vec{G}',
\vec{K}) = 0, \label{eq:final03}\EE with
$$Q_{\vec{G},\vec{G}'}(\vec{K},\omega) =
[(\vec{G}+\vec{K})\cdot(\vec{G}'+\vec{K})]A(\vec{G}-\vec{G}') -
\frac{\omega^2}{g}\delta_{\vec{G}, \vec{G}'}.$$ The dispersion
relation connecting $\vec{K}$ and $\omega$ is determined by the
secular equation \BE \mbox{det}\left[
(\vec{G}+\vec{K})\cdot(\vec{G}'+\vec{K})]A(\vec{G}-\vec{G}') -
\frac{\omega^2}{g}\delta_{\vec{G},
\vec{G}'}\right]_{\vec{G},\vec{G}'} = 0.\EE

For the shallow water, we have $\tanh(kh) \approx kh$, and thus
$\tanh(kh)/k \approx h$, then by \BE h(x,y) = \sum_{\vec{G}}
A(\vec{G})e^{i\vec{G}\cdot\vec{r}},\label{eq:h1}\EE with \BE
A(\vec{G}) = \left\{ \begin{array}{ll} (h_1-h) f_s + h, &
\mbox{for} \ \vec{G}=0;\\ (h_1 - h)F_s(\vec{G}), & \mbox{for} \
\vec{G} \neq 0.\end{array}\right.\EE

\subsection{Multiple scattering theory}

The shallow water wave propagation in the water column with
cylindrical steps can also be investigated by the multiple
scattering theory. Without requiring that all the steps are the
same, we can develop a general formulism.

In the water column, the wave equation reads \BE (\nabla^2 +
k^2)\eta = 0, \label{eq:wave01}\EE with $k$ being given by
$$\omega^2 = gk\tanh(kh)$$ Within the range of the $i$-th
step, the wave equation is \BE (\nabla^2 + k_i^2)\eta_i = 0,
\label{eq:wave02}\EE with $$ \omega^2 = gk\tanh(kh)$$ At the
boundary of the step, the boundary conditions are \BE \left.
\frac{\tanh(k_ih_i)}{k_i} \hat{n}\cdot\nabla\eta_i
\right|_{\partial \Omega_i} = \left.
\frac{\tanh(k_ih_i)}{k_i}\hat{n}\cdot\nabla\eta \right|_{\partial
\Omega_i}, \label{eq:bc1}\EE derived from the conservation of
mass, and \BE \label{eq:bc2} \left.\eta_i\right|_{\partial
\Omega_i} = \left.\eta\right|_{\partial \Omega_i}.\EE Here
$\partial\Omega_i$ denotes the boundary, and $\hat{n}$ is the
outward normal at the boundary.

Equations (\ref{eq:wave01}) and (\ref{eq:wave02}) with the
boundary conditions in (\ref{eq:bc1}) and (\ref{eq:bc2})
completely determine the shallow water wave scattering by an
ensemble of cylindrical steps located vertically in the uniform
water column. By inspecting, we see that this set of equations is
essentially the same as the two dimensional acoustic scattering by
an array of parallel cylinders\cite{Twersky,Chen}. We following
\cite{Chen} to study the scattering of shallow water waves in the
present system.

Consider a line source located at $\vec{r}_s$. Without the
cylinder steps, the wave is governed by \BE (\nabla^2
+k^2)G(\vec{r}-\vec{r}_s) = -4\pi\delta^{(2)}(\vec{r} -
\vec{r}_s), \EE where $H_0^{(1)}$ is the zero-th order Hankel
function of the first kind. In the cylindrical coordinates, the
solution is \BE G(\vec{r}-\vec{r}_s) = \mbox{i}\pi
H_0^{(1)}(k|\vec{r} - \vec{r}_s|). \EE In this section,
`$\mbox{i}$' stands for $\sqrt{-1}$.

With $N$ cylinder steps located at $\vec{r}_i$ ($i=1,2,\cdots,
N$), the scattered wave from the $j$-th step can be written as \BE
\label{eqps1} \eta_s(\vec{r}, \vec{r}_j) =
\sum_{n=-\infty}^{\infty} \mbox{i}\pi A_n^j H_n^{(1)}(k|\vec{r} -
\vec{r}_j|)e^{\mbox{i}n\phi_{\vec{r}- \vec{r}_j}}, \EE where
$H_n^{(1)}$ is the $n$-th order Hankel function of the first kind.
$A_n^i$ is the coefficient to be determined, and $\phi_{\vec{r}-
\vec{r}_j}$ is the azimuthal angle of the vector $\vec{r}-
\vec{r}_i$ relative to the positive $x$-axis.

The total wave incident around the $i$-th scatterer
$\eta_{in}^i(\vec{r})$ is a superposition of the direct
contribution from the source $\eta_0(\vec{r}) =
G(\vec{r}-\vec{r}_s)$ and the scattered waves from all other
scatterers: \BE \label{eqpin1} \eta_{in}^i(\vec{r}) =
\eta_0(\vec{r}) + \sum_{j=1,j\neq i}^N \eta_s(\vec{r}, \vec{r}_j).
\EE In order to separate the governing equations into modes, we
can express the total incident wave in term of the modes about
$\vec{r_i}$: \BE \label{eqpin2} \eta_{in}^i(\vec{r}) = \sum_{n =
-\infty}^\infty B_n^i J_n(k|\vec{r} -
\vec{r_i}|)e^{\mbox{i}n\phi_{\vec{r} - \vec{r_i}}}. \EE The
expansion is in terms of Bessel functions of the first kind $J_n$
to ensure that $\eta_{in}^i(\vec{r})$ does not diverge as $\vec{r}
\rightarrow \vec{r_i}$.  The coefficients $B_n^i$ are related to
the $A_n^j$ in equation (\ref{eqps1}) through equation
(\ref{eqpin1}).  A particular $B_n^i$ represents the strength of
the $n$-th mode of the total incident wave on the $i$-th scatterer
with respect to the $i$-th scatterer's coordinate system (i.e.
around $\vec{r_i}$).  In order to isolate this mode on the right
hand side of equation (\ref{eqpin1}), and thus determine a
particular $B_n^i$ in terms of the set of $A_n^j$, we need to
express $\eta_s(\vec{r}, \vec{r_j})$, for each $j \neq i$, in
terms of the modes with respect to the $i$-th scatterer. In other
words, we want $\eta_s(\vec{r}, \vec{r_j})$ in the form \BE
\label{eqps2} \eta_s(\vec{r}, \vec{r_j}) = \sum_{n=-\infty}^\infty
C_n^{j, i} J_n(k|\vec{r} - \vec{r_i}|)e^{\mbox{i}\phi_{\vec{r} -
\vec{r_i}}}. \EE This can be achieved (i.e. $C_n^{j, i}$ expressed
in terms of $A_n^i$) through the following addition
theorem\cite{addition}: \BA \label{eqaddition} H_n^{(1)}(k|\vec{r}
- \vec{r_j}|)e^{\mbox{i}n\phi_{\vec{r} - \vec{r_j}}} =
e^{\mbox{i}n\phi_{\vec{r_i} - \vec{r_j}}}\times && \nonumber\\
\sum_{l=-\infty}^{\infty} H_{n-l}^{(1)}(k|\vec{r_i} - \vec{r_j}|)
e^{-\mbox{i}l\phi_{\vec{r_i} - \vec{r_j}}} J_l(k|\vec{r} -
\vec{r_i}|) e^{\mbox{i}l\phi_{\vec{r} - \vec{r_i}}}. && \EA

Taking equation (\ref{eqaddition}) into equation (\ref{eqps1}), we
have \BA \eta_s(\vec{r}, \vec{r_j}) = \sum_{n=-\infty}^\infty
\mbox{i}\pi
A_n^j e^{\mbox{i}n\phi_{\vec{r_i} - \vec{r_j}}} && \nonumber\\
\sum_{l=-\infty}^{\infty} H_{n-l}^{(1)}(k|\vec{r_i} - \vec{r_j}|)
e^{-\mbox{i}l\phi_{\vec{r_i} - \vec{r_j}}} J_l(k|\vec{r} -
\vec{r_i}|) e^{\mbox{i}l\phi_{\vec{r} - \vec{r_i}}}. && \EA
Comparing with equation (\ref{eqps2}), we see that \BE C_n^{j,i} =
\sum_{l=-\infty}^\infty \mbox{i}\pi A_l^j
H_{l-n}^{(1)}(k|\vec{r_i} - \vec{r_j}|)
e^{\mbox{i}(l-n)\phi_{\vec{r_i} - \vec{r_j}}} \EE

Now we can relate $B_n^i$ to $C_n^{j, i}$ (and thus to $A_l^j$)
through equation (\ref{eqpin1}).  First note that through the
addition theorem the source wave can be written, \BE
\label{eqp0exp}
\begin{array}{lll}
\eta_0(\vec{r}) & = & \mbox{i}\pi H_0^{(1)}(k|\vec{r}-\vec{r}_s|) \\
& = & \sum_{l=-\infty}^{\infty} S_l^i J_l(k|\vec{r} - \vec{r_i}|)
e^{\mbox{i}l\phi_{\vec{r} - \vec{r_i}}},
\end{array}
\EE where \BE S_l^i = \mbox{i}\pi
H_{-l}^{(1)}(k|\vec{r_i}-\vec{r}_s|)
e^{-\mbox{i}l\phi_{\vec{r_i}}}. \EE Matching coefficients in
equation (\ref{eqpin1}) and using equations (\ref{eqpin2}),
(\ref{eqps2}) and (\ref{eqp0exp}), we have \BE B_n^i = S_n^i +
\sum_{j=1,j\neq i}^N C_n^{j, i}, \EE or, expanding $C_n^{j, i}$,
\BE \label{eqmatrix1} B_n^i = S_n^i + \sum_{j=1,j\neq i}^N
\sum_{l=-\infty}^\infty \mbox{i}\pi A_l^j
H_{l-n}^{(1)}(k|\vec{r_i} - \vec{r_j}|)
e^{\mbox{i}(l-n)\phi_{\vec{r_i} - \vec{r_j}}}. \EE At this stage,
both the $S_n^i$ are known, but both $B_n^i$ and $A_l^j$ are
unknown.  Boundary conditions will give another equation relating
them.

The wave inside the $i$-th scatterer can be expressed as \BE
\label{eqpint1} \eta_{int}^i(\vec{r}) = \sum_{n = -\infty}^\infty
D_n^i J_n(k_i|\vec{r} - \vec{r_i}|) e^{\mbox{i}n\phi_{\vec{r} -
\vec{r_i}}}. \EE Taking Eqs.~(\ref{eqps1}), (\ref{eqpin2}), and
(\ref{eqpint1}) into the boundary conditions in (\ref{eq:bc1}) and
(\ref{eq:bc2}), we have \BA B_n^i J_n(k a_i) + \mbox{i}\pi A_n^i
H_n^{(1)}(k a_i) & = & D_n^i J_n(k_i a_i) \\ B_n^i J_n'(k a_i) +
\mbox{i}\pi A_n^i H_n^{(1)\prime}(k a_i) & = &
\frac{\tanh(h_ik_i)}{\tanh(hk)} D_n^i J_n'(k_i a_i),
\nonumber\\\EA where `${^\prime}$' refers to the derivative.
Elimination of $D_n^i$ gives \BE B_n^i = \mbox{i}\pi\Gamma_n^i
A_n^i, \EE where \BE \label{eq:Gamma1}\Gamma_n^i =
\frac{H_n^{(1)}(k a_i) J_n'(k_i a_i) -
\frac{\tanh(kh)}{\tanh(k_ih_i)} H_n^{(1)\prime}(k a_i) J_n(k_i
a_i)} {\frac{\tanh(kh)}{\tanh(k_ih_i)} J_n'(k a_i) J_n(k_i a_i) -
J_n(k a_i) J_n'(k_i a_i)}. \EE If we define \BE T_n^i =
S_n^i/\mbox{i}\pi = H_{-n}^{(1)}(k|\vec{r_i}-\vec{r}_s|)
e^{-\mbox{i}n\phi_{\vec{r_i}}} \EE and \BE G_{l,n}^{i,j} =
H_{l-n}^{(1)}(k|\vec{r_i} - \vec{r_j}|)
e^{\mbox{i}(l-n)\phi_{\vec{r_i} - \vec{r_j}}}, i\neq j \EE then
equation (\ref{eqmatrix1}) becomes \BE \label{eqfinalmatrix}
\Gamma_n^i A_n^i - \sum_{j=1,j\neq i}^N \sum_{l=-\infty}^\infty
G_{l,n}^{i,j} A_l^j = T_n^i. \EE If the value of $n$ is limited to
some finite range, then this is a matrix equation for the
coefficients $A_n^i$.  Once solved, the total wave at any point
outside all cylinder steps is \BA \eta(\vec{r}) &=& \mbox{i}\pi
H_0^{(1)}(k|\vec{r}-\vec{r}_s|) + \nonumber \\
& & \sum_{i=1}^N \sum_{n=-\infty}^\infty \mbox{i}\pi A_n^i
H_n^{(1)}(k|\vec{r} - \vec{r_i}|) e^{\mbox{i}n\phi_{\vec{r} -
\vec{r_i}}}.\label{eq:final} \EA We must stress that total wave
expressed by eq.~(\ref{eq:final}) incorporate all orders of
multiple scattering. We also emphasize that the above derivation
is valid for any configuration of the cylinder steps. In other
words, eq.~(\ref{eq:final}) works for situations that the steps
can be placed either randomly or orderly.

For the special case of shallow water ($kh <<1$), we need just
replace $\Gamma_n^i$ in Eq.~(\ref{eq:Gamma1}) by \BE
\label{eq:Gamma2}\Gamma_n^i = \frac{H_n^{(1)}(k a_i) J_n'(k_i a_i)
- \sqrt{\frac{h}{h_i}} H_n^{(1)\prime}(k a_i) J_n(k_i a_i)}
{\sqrt{\frac{h}{h_i}} J_n'(k a_i) J_n(k_i a_i) - J_n(k a_i)
J_n'(k_i a_i)}. \EE

\section{Summary}

In summary, here we have presented a general theory for studying
gravity waves over bottom topographies. The results have been
extended to the case of step-wise bottom structures. The model
presented here is simple and may facilitate the research on many
unusual wave phenomena such as wave localization\cite{Im,Emile}.

\begin{center}
 {\bf Acknowledgments}
\end{center}
Discussion with H.-P. Fang and X.-H. Hu at Fudan University are
appreciated. The comments from X.-H. Hu are acknowledged. The
helps from K.-H. Wang, B.~Gupta, and P.-C. (Betsy) Cheng are also
thanked.

\end{document}